 \documentclass[12pt,preprint,preprint,nofootinbib]{revtex4}
 \pdfoutput=1
\usepackage{epsf, array, color}
\usepackage{graphicx}
\usepackage{subfigure}
\usepackage{bbold}
\usepackage{amsmath}
\usepackage{mathtools}
\usepackage{dcolumn}
\usepackage{url}
\usepackage{verbatim}
\usepackage{wrapfig}
\usepackage{amsfonts}
\usepackage{comment}
\usepackage{epigraph}
\usepackage{slashed}
\usepackage[utf8]{inputenc}
\usepackage[linktocpage,colorlinks,urlcolor=blue]{hyperref}

\usepackage{bbm}
\usepackage{cancel}
\usepackage{epsf, array, color}

\newcommand{\nc}{\newcommand}

\newcommand{\bea}{\begin{eqnarray}}
\newcommand{\eea}{\end{eqnarray}}
\newcommand{\nn}{\nonumber \\}

\def\beq{\begin{equation}}
\def\eeq{\end{equation}}

\newcommand{\co}{\mathcal{O}}

\newcommand{\C}{\mathcal{C}}

\newcommand{\Op}{{\cal O}}
\nc{\vp}{\phi}
\nc{\tvp}{\widetilde{\phi}}
\nc{\vpj }{\mbox{${\vp^\dag i\,\raisebox{2mm}{\boldmath ${}^\leftrightarrow$}\hspace{-4mm} D_\mu\,\vp}$}}
\nc{\vpjt}{\mbox{${\vp^\dag i\,\raisebox{2mm}{\boldmath ${}^\leftrightarrow$}\hspace{-4mm} D_\mu^{\,I}\,\vp}$}}

\def\nn{\nonumber}
\def\gev{\rm GeV}

\begin{document}
\title{Electroweak Corrections to  Higgs to $\gamma\gamma$ and $W^+W^-$ in the SMEFT}

\author{S.~Dawson and P. P. Giardino  }
\affiliation{
\vspace*{.5cm}
  \mbox{Department of Physics,\\
  Brookhaven National Laboratory, Upton, N.Y., 11973,  U.S.A.}\\
 \vspace*{1cm}}

\date{\today}

\begin{abstract}
Higgs decays to gauge boson pairs are a crucial ingredient in the study of Higgs properties, with   the decay $H\rightarrow\gamma\gamma$  being particularly sensitive to new physics effects.  Assuming all potential new physics occurs at energies much above the weak scale, deviations from Standard Model predictions can be parameterized in terms of the coefficients of an effective field theory (SMEFT).  When experimental limits on the SMEFT coefficients reach an accuracy of a few percent, predictions must be done beyond the lowest order in  the SMEFT in order to match theory and experimental accuracy.  This paper completes a program of computing the one-loop electroweak SMEFT corrections to $H\rightarrow VV^\prime$, $V=W^\pm,Z,\gamma$.  The  calculation of the real contribution to $H\rightarrow W^+W^-\gamma$ is performed by mapping two-loop amplitudes to the $3-$ body phase space.%
\end{abstract}
\maketitle

\section{Introduction}

The LHC Higgs program is entering an era of precision measurements that requires a program of higher
order theoretical calculations .  The need for  precise calculations is driven by:
$(1)$ the non-discovery of new particles that implies that the scale of Beyond the Standard Model  (BSM) physics 
must typically be  much higher than $\sim 1~TeV$ and
$(2)$  the anticipated precision in the Higgs measurements
at the high luminosity LHC  \cite{CMS-PAS-FTR-16-002,ATL-PHYS-PUB-2013-014}.  
In order to study  deviations of Higgs properties from the SM predictions,  a consistent theoretical framework is needed so
that the accuracy of theoretical calculations is comparable to that of the measurements.  

The Standard Model (SM)  QCD  and electroweak contributions to Higgs production and decay
are known to at least NLO for all relevant processes
and provide a framework for comparison  \cite{deFlorian:2016spz}.  In the LHC Run-1, deviations of Higgs measurements from SM predictions were typically expressed in 
terms of limits on coupling constant modifiers  \cite{Khachatryan:2016vau}.   This $\kappa$ approach rescales all Higgs couplings by constant factors and is not sensitive to kinematic distributions. 
  As measurements approach the level of $5-10\%$ accuracy, however, it becomes
necessary to include electroweak corrections to the predictions, which in turn necessitates the use of effective field theory techniques, since
electroweak corrections typically cannot be incorporated into a simple  rescaling of the Higgs couplings.  

The use of effective field theories for studying Higgs production and decay is well established  \cite{Giudice:2007fh,Brivio:2017vri,Contino:2014aaa}.  The
SM effective field theory  (SMEFT)  assumes that the Higgs is an $SU(2)_L$ doublet and parameterizes new physics through an expansion in higher dimensional operators,
\begin{equation}
{\cal L}={\cal L}_{SM}+\Sigma_{k=5}^{\infty}\Sigma_{i=1}^n {\mathcal{C}_i^k\over \Lambda^{k-4}} O_i^k\, ,
\label{eq:lsmeft}
\end{equation}
where the $SU(3)\times SU(2)_L\times U(1)_Y$ invariant 
dimension-$k$ operators are constructed from SM fields and all of  the
 BSM physics effects reside in the coefficient functions, $\C_i^k$.    If the scale $\Lambda>>v$, then it suffices to truncate
 the expansion at dimension-6.  
 This large separation of scales is necessary in order for a study containing  only dimension-6 operators be
 valid, since the effects of the dimension-8 operators are assumed to be suppressed by an additional factor of $v^2/\Lambda^2$ and are neglected.  Similarly, it is assumed that there are no new particles in the theory at scales below $\Lambda$.
 
 We need predictions to NLO QCD and EW accuracy in the SMEFT so that the theoretical 
 predictions have roughly the same uncertainties as the experimental results.  
For processes with strong interactions, many NLO
QCD results in the SMEFT exist, particularly in the top-Higgs sector \cite{AguilarSaavedra:2018nen}.  Electroweak corrections
in the SMEFT \cite{Ghezzi:2015vva} are available for only a handful of processes: $H\rightarrow b {\overline b}$ \cite{Gauld:2016kuu,Gauld:2015lmb},
 $H\rightarrow \gamma \gamma$ \cite{Hartmann:2015aia,Hartmann:2015oia,Dedes:2018seb,Ghezzi:2015vva}, $H\rightarrow Z\gamma$ \cite{Dawson:2018pyl} and $Z\rightarrow
 f{\overline f}$ \cite{Hartmann:2016pil}.  Here, we complete the program of computing 
 the on-shell decays $H\rightarrow VV^\prime$,
($V=Z,W^\pm,\gamma$), at one-loop in the SMEFT.  
Previously, we presented one-loop SMEFT results for  $H\rightarrow Z\gamma$ and for the (unphysical) on-shell decay $H\rightarrow ZZ$ and  \cite{Dawson:2018pyl}. 

In this paper, we present the one-loop SMEFT results for $H\rightarrow \gamma\gamma$ and  the on-shell process $H\rightarrow W^+W^-$.  The result for the 
decay $H\rightarrow\gamma
\gamma$
follows from the results of Ref.  \cite{Dawson:2018pyl} and we compare with the results\footnote{The corrections due to the top-loops in the SMEFT have been studied also in \cite{Vryonidou:2018eyv}.} of Refs.  \cite{Hartmann:2015aia,Hartmann:2015oia,Dedes:2018seb}.  We consider two different input parameter choices in order
to assess their numerical significance.  
Our result contains the full (constant plus logarithmic terms) SMEFT result for the renormalization of
 $G_F$\footnote{ Refs.  \cite{Hartmann:2015aia,Hartmann:2015oia,Ghezzi:2015vva} contain only the logarithmic
 contributions to the SMEFT renormalization of $G_F$. }.
  Our one-loop $H\rightarrow W^+W^-$ result is an intermediate step on the way to the physical process $H\rightarrow W^+W^-\rightarrow$ 4 fermions. 
 The  calculation  of the real contributions from $H\rightarrow W^+W^-\gamma$ is performed using a 
 mapping of the 3-body phase space to 2-loop amplitudes, which is of technical interest \cite{Anastasiou:2002yz}.

Section \ref{sec:basics} reviews the one-loop electroweak renormalization for $H\rightarrow VV$ decays, Section \ref{sec:wwres} has results for $H\rightarrow W^+W^-$,
and Section \ref{sec:ggsec} contains the one-loop results for $H\rightarrow \gamma\gamma$.  Conclusions are contained in Section \ref{sec:conc}.

\section{Basics}
\label{sec:basics}

We use the Warsaw basis \cite{Buchmuller:1985jz,Grzadkowski:2010es} where the relevant operators for the one-loop contributions
to the decays $H\rightarrow VV$ are given in Table \ref{tab:opdef}  and the Feynman rules and
conventions in $R_\xi$ gauge are taken from Ref.  \cite{Dedes:2017zog}. 
\begin{table}[t] 
\centering
\renewcommand{\arraystretch}{1.5}
\begin{tabular}{||c|c||c|c||c|c||} 
\hline \hline
$\Op_W$                & $\epsilon^{IJK} W_\mu^{I\nu} W_\nu^{J\rho} W_\rho^{K\mu}$ &    
 $\Op_{\vp\Box}$ & $(\vp^\dag \vp)\raisebox{-.5mm}{$\Box$}(\vp^\dag \vp)$ 
 &
$\Op_{\vp D}$   & $\left(\vp^\dag D^\mu\vp\right)^* \left(\vp^\dag D_\mu\vp\right)$ 
\\
\hline 
$\underset{p,r}{\Op_{u\vp}}$           & $(\vp^\dag \vp)(\bar q'_p u'_r \tvp)$&
 $\Op_{\vp W}$    & $ (\vp^\dag \vp)\, W_{\mu\nu} W^{\mu\nu}$ 
   &
   $\Op_{\vp B}$     & $ (\vp^\dag \vp)\, B_{\mu\nu} B^{\mu\nu}$
  \\ \hline 
     $\Op_{\vp WB}$     & $ (\vp^\dag \tau^I \vp)\, W^I_{\mu\nu} B^{\mu\nu}$ &
$\Op_{uW}$               & $(\bar q'_p \sigma^{\mu\nu} u'_r) \tau^I \tvp\, W_{\mu\nu}^I$ & 
$\underset{p,r}{\Op_{uB}}$        & $(\bar q'_p \sigma^{\mu\nu} u'_r) \tvp\, B_{\mu\nu}$
\\
\hline
 $\underset{p,r}{\Op_{\vp l}^{(3)}}$      & $(\vpjt)(\bar l'_p \tau^I \gamma^\mu l'_r)$
&
    $\underset{p,r,s,t}{\Op_{ll}}$        & $(\bar l'_p \gamma_\mu l'_r)(\bar l'_s \gamma^\mu l'_t)$  
    &&

\\
\hline \hline
\end{tabular}
\caption{Dimension-6 operators relevant for the one-loop contributions to $H\rightarrow VV$
($V=W,Z,\gamma$  (from  \cite{Grzadkowski:2010es}). For brevity we suppress fermion
  chiral indices $L,R$. $I=1,2,3$ is an $SU(2)$ index, $p,r$ are flavor
  indices,  and $\vpj\equiv \phi^\dagger D_\mu \phi- (D_\mu\phi^\dagger) \phi$.  \label{tab:opdef}}
\end{table}
For simplicity, we assume a diagonal flavor structure for the coefficients $\mathcal{C}$, {\it i.e.} $\underset{p,r}{\C_i}=\C_i \underset{p,r}{\mathbb{1}}$,
where $p,r$ are flavor indices. 
 Furthermore, we assume $\underset{e,\mu,\mu,e}{\C_{ll}}=\underset{\mu,e,e\mu}{\C_{ll}}\equiv \C_{ll}$ and 
 $\underset{\mu,\mu,t,t}{\C_{lq}^{(3)}}=\underset{e,e,t,t}{\C_{lq}^{(3)}}\equiv \C_{lq}^{(3)}$.

The Higgs Lagrangian is,
\bea
\mathcal{L}&=&(D_\mu \phi )^\dagger (D_\mu \phi )+ \mu ^2 \phi ^\dagger \phi  -\lambda (\phi ^\dagger \phi )^2 \nn\\
&+& \frac1{\Lambda^2} \biggl(
\C_\phi  (\phi ^\dagger \phi )^3+\C_{\phi \square}  (\phi ^\dagger \phi )\square(\phi ^\dagger \phi )
+\C_{\phi D}(\phi ^\dagger D_\mu \phi)^*(\phi^\dagger D_\mu\phi)\biggr)\, ,
\label{eq:HiggsPotential}
\eea
where  $\phi$  is the usual Higgs doublet:
\bea
\phi =\left(
\begin{array}{c}
\phi^+ \\
\frac1{\sqrt{2}}(v+H+i \phi^0)
\end{array}
\right),
\eea
and $v$ is the vacuum expectation value (vev) defined as the minimum of the potential, 
\beq
v\equiv\sqrt{2}\langle \phi \rangle= \sqrt\frac{\mu^2}{\lambda}+\frac{3 \mu^3}{8 \lambda^{5/2}}\frac{\C_\phi }{\Lambda^2}.
\eeq
The Higgs kinetic terms in the resulting Lagrangian are not canonically normalized due to
$\co_{\phi \square}$ and $\co_{\phi D}$.
As a consequence we need to shift the fields, 
\bea
H&\rightarrow & H\biggl(1-\frac{v^2}{\Lambda^2}(\frac14 \C_{\phi D}-\C_{\phi \square})\biggr) \nonumber\\
\phi^0&\rightarrow&\phi^0\biggl(1-\frac{v^2}{\Lambda^2}(\frac14 \C_{\phi D})\biggr) \nonumber\\
\phi^+&\rightarrow&\phi^+\, .
\label{eq:HiggsShift}
\eea
The physical mass of the Higgs to ${\cal{O}}\biggl({1\over\Lambda^2}\biggr)$ becomes,
\beq
M_H^2=2\lambda v^2-\frac{v^4}{\Lambda^2}(3 \C_\phi -4 \lambda \C_{\phi \square}+\lambda \C_{\phi D}).
\eeq

The SMEFT interactions  also cause the gauge field kinetic energies to have non-canonical normalizations and
following Ref.  \cite{Dedes:2017zog}, we define "barred" fields and couplings,
\begin{eqnarray}
{\overline W}_\mu & \equiv & (1-\C_{\phi W} v^2/\Lambda^2)W_\mu
\nonumber \\
{\overline B}_\mu & \equiv & (1-\C_{\phi B}v^2/\Lambda^2)B_\mu
\nonumber \\
{\overline g}_2 & \equiv &(1+\C_{\phi W} v^2/\Lambda^2)g_2
\nonumber \\
{\overline g}_1 & \equiv&  (1+\C_{\phi B}v^2/\Lambda^2)g_1
\end{eqnarray}
such  that ${\overline W}_\mu {\overline g}_2= W_\mu g_2$ and ${\overline B}_\mu {\overline g}_1= B_\mu g_1$. 
The "barred" fields defined in this way have  properly normalized  kinetic  energy terms. 
The masses of the W and Z fields to ${\cal {O}}\biggl({1\over \Lambda^2}\biggr)$ are
  \cite{Dedes:2017zog,Alonso:2013hga},
\bea
M_W^2&=&\frac{{\overline g}_2^2 v^2}4,\nn\\
M_Z^2&=&\frac{({\overline g}_1^2+{\overline g}_2^2) v^2}4+\frac{v^4}{\Lambda^2}\left(\frac18 ({\overline g}_1^2+{\overline g}_2^2) \C_{\phi D}+\frac12 {\overline g}_1{\overline g}_2\C_{\phi WB} \right).
\eea

Dimension-6 4-fermion operators    give contributions to the decay of the $\mu$, changing the relation between the 
vev, $v$, and the Fermi constant $G_\mu$.  Considering only contributions that interfere with the
 SM amplitude,  we obtain the  tree level result,
\begin{eqnarray}
G_\mu
\equiv \frac1{\sqrt{2} v^2}-\frac1{\sqrt{2}\Lambda^2}\C_{ll}+{\sqrt{2}\over \Lambda^2}\C_{\phi l}^{(3)},
\label{eq:gdef}
\end{eqnarray}
where we assume the $\C_i$ are flavor universal.
The tadpole counterterms are defined such that they cancel completely the tadpole graphs  \cite{Fleischer:1980ub}. 
 This condition identifies the renormalized vacuum as the minimum of the renormalized scalar potential
 at each order of perturbation theory.

 Since the SMEFT theory is only renormalizable order by order in the $(v^2/\Lambda^2)$ expansion, 
we  drop all  terms proportional to $(v^2/\Lambda^2)^a$ with $a>1$.
The one-loop SMEFT calculations contain  both tree level
and one-loop
contributions from the dimension-6 operators, along with the  full electroweak one-loop SM amplitudes.

We use  a modified on shell (OS) scheme, where the SM parameters are OS quantities.
Since the coefficients of the dimension-6 operators are not physical observables, we  treat them as $\overline{{MS}}$
parameters, so the renormalized coefficients are  $\C(\mu)=\C_0-\text{poles}$, where $\C_0$ are the bare quantities. 
The poles of the coefficients $\C_0$ are found from the renormalization group (RG) evolution of the coefficients computed in the unbroken
phase of the theory in 
Refs.  \cite{Jenkins:2013zja,Jenkins:2013wua,Alonso:2013hga},
\beq
\C_i(\mu)=\C_{0,i}-\frac1{2\hat{\epsilon}}\frac1{16\pi^2}\gamma_{ij}\C_j,
\eeq
where $\mu$ is the renormalization scale,  $\gamma_{ij}$ is the one-loop anomalous dimension, 
\beq
\mu \frac{d \C_i}{d\mu}=\frac1{16\pi^2}\gamma_{ij}\C_j,
\eeq
and $\hat{\epsilon}^{-1}\equiv\epsilon^{-1}-\gamma_E+\log(4\pi)$.
At one-loop,  tree level parameters (denoted with the subscript 0 in this section) must be renormalized.
The renormalized SM masses are defined by,
\beq
M_V^2=M^2_{0,V}-\Pi_{VV}(M^2_{V}),
\eeq
where $\Pi_{VV}(M^2_V)$ is the one-loop correction to the 2-point function for Z or W computed on-shell.
The gauge boson $2$- point functions in the SMEFT can be found  analytically in Refs.  \cite{Chen:2013kfa,Ghezzi:2015vva}.

The one- loop relation between the vev and the Fermi constant is,
\beq
G_\mu+{\C_{ll}\over\sqrt{2}\Lambda^2}-\sqrt{2}{\C_{\phi l}^{(3)}\over\Lambda^2}\equiv\frac1{\sqrt{2} v_0^2}(1+\Delta r),
\label{eq:geftdef}
\eeq
where $v_0$ is the unrenormalized minimum of the potential and
 $\Delta r$ is obtained from the one-loop corrections to $\mu$ decay.
 Analytic expressions for $\Delta r$ in both the SM and the SMEFT at dimension-$6$ are given in Ref.  \cite{Dawson:2018pyl}. 
 
The calculation proceeds in the same way as Ref.  \cite{Dawson:2018pyl}. We obtain the relevant amplitudes using FeynArts  \cite{Hahn:2000kx} with a model file generated by FeynRules \cite{Alloul:2013bka} with the Feynman rules presented in \cite{Dedes:2017zog}. Then we use FeynCalc \cite{Mertig:1990an,Shtabovenko:2016sxi} to manipulate and reduce the integrals and LoopTools \cite{Hahn:2000jm} for the numerical evaluation.

We consider two choices of input parameters.  For the $W^+W^-$ calculation,
we choose the $G_\mu$ scheme, where we take the physical  input parameters to be
\footnote{
The light quark masses and lepton masses enter
into the $\gamma$ wave-function renormalization for $H\rightarrow \gamma \gamma$ and we take $m_b=4.78~\gev,\, m_c=1.67~\gev,\, 
m_s=0.1~\gev,\, m_d=0.005~\gev,\, m_u=0.002~\gev,\, m_\tau=1.776~\gev,\, m_\mu=0.105~\gev\,\text{and}\, m_e=0.0005~\gev$.}
 \begin{eqnarray}
G_\mu&=&1.1663787(6)\times 10^{-5} \gev^{-2}\nonumber \\
M_Z&=&91.1876\pm .0021\gev\nonumber \\
M_W&=&80.385\pm .015~\gev\nonumber\\
M_H&=&125.09\pm 0.21\pm 0.11 ~\gev\nonumber\\
M_t&=&173.1\pm0.6~\gev\nonumber\, .
\label{eq:inputs}
\end{eqnarray}
We then follow the same procedure as in Ref.  \cite{Dawson:2018pyl}.  In our discussion, we term this the "$G_\mu,~M_W,~M_Z$ scheme".

 For  the decay $H\rightarrow \gamma\gamma$,  
 we consider the effects of explicitly pulling out an overall factor of $\alpha$ from the amplitudes, that is we calculate
 \begin{equation}
 \mathcal{A}(H\to\gamma\gamma)=\alpha_0{{\hat{F}}}(v_0,\, M_{0,W},\, M_{0,Z}),
 \end{equation}
 where ${F}$ is a function of the bare parameters $v_0,\,M_{0,W},\,M_{0,Z},$ that we renormalize as described before and express in terms of $G_\mu$, $M_W$ and $M_Z$. The on-shell renormalization of the overall factor $\alpha$ is extracted from the renormalization of the $\gamma\bar{l} l$ vertex and we take the physical parameter 
  \begin{equation}
 \alpha=\frac1{137.035999139(31)}.
 \end{equation}
 We term this the "$\alpha, G_\mu,~M_W,~M_Z$  scheme".

\section{$H\rightarrow W^+W^-$}
\label{sec:wwres}

The tree level decay width for $H\rightarrow W^+W^-$  receives contributions from the rescaling of the Higgs field (Eq. \ref{eq:HiggsShift}), 
the SMEFT 
contributions fo $G_\mu$ (Eq. \ref{eq:gdef}) and the direct interaction of $O_{\phi W}.$
For $M_H=200~GeV$, the numerical result SMEFT tree level result  in GeV  is, 
\begin{eqnarray}
\Gamma_{0}(H\rightarrow W^+W^-)&=&1.042+  \biggl({1~TeV\over \Lambda}\biggr)^2\biggl\{
0.1263 \biggl( \C_{ \phi \square}- \C_{\phi l}^{(3)}-{1\over 4} C_{\phi D}\biggr)
\nonumber \\ && -0.2485\biggl(\C_{\phi W}
-0.2541\C_{ll}\biggr)\biggr\}
\nonumber \\ && 
+\biggl({1~TeV\over \Lambda}\biggr)^4\biggl\{
  0.003828 \biggl( \C_{ \phi \square}- \C_{\phi l}^{(3)}-{1\over 4} C_{\phi D}\biggr)^2
 \ \nonumber \\ && 
- 0.01506 \biggl( \C_{\phi W}-.2541\C_{ll}\biggr)              \biggl( \C_{\phi \square}
 -\C_{ \phi l}^{(3)}-{1\over 4}  \C_{\phi D}  \biggr)  
  \nonumber \\ && 
 + 0.02343 \C_{\phi W}^2   
- 
 0.007531 \C_{\phi W} \C_{ll} 
 +0.0009570\C_{ll}^2\biggr\}\, . 
\label{eq:loww}
\end{eqnarray}
We have retained terms of ${\cal {O}}(\C_i^2)$ in Eq. \ref{eq:loww}  although the
 numerical  coefficients are suppressed relative to those of the  ${\cal {O}}(\C_i)$ terms.
 The usual tree level scaling factor is defined to ${\cal{O}}({1\over\Lambda^2})$,
 \begin{eqnarray}
 \mu_0(H\rightarrow W^+W^-)&=&{\Gamma_0(H\rightarrow W^+W^-)\over \Gamma_0(H\rightarrow W^+W^-)\mid_{SM}}
 \nonumber \\
 &\rightarrow &1.0 
 + \biggl({1~TeV\over \Lambda}\biggr)^2\biggl\{0.1212\biggl(\C_{\phi\square}-  \C_{\phi l}^{(3)} 
 -{1\over 4} \C_{\phi D}\biggr) 
\nonumber \\ && 
 - 0.2385 \C_{\phi W} +0.06060 \C_{ll} \biggr\},\quad  {\text{for}}~M_H=200~GeV\, .
 \end{eqnarray}
 
 At NLO in the SMEFT, the decay width receives contributions from the one-loop virtual diagrams,
 including the renormalization terms discussed in the previous section, and from real photon emission.  These contributions are 
 separately IR divergent and
 we regulate them with a photon mass. 
  The SM rate including all electroweak
 corrections is well known, both for the on-shell decay $H\rightarrow W^+W^-$  \cite{Kniehl:1991xe}
 and the off-shell decays, $H\rightarrow $4 fermions \cite{Bredenstein:2006ha}. 
 The off-shell effects are known to be significant for the physical $M_H=125~GeV$ Higgs and the extension of our calculation to include the off-shell effects is clearly a needed step. 
 The SM
 electroweak corrections are of order $\sim6\%$ for our reference Higgs mass, $M_H=200~GeV$.
 
 The calculation of the virtual contribution in the SMEFT  follows the identical procedure as for $H\rightarrow ZZ$, with
 the exception of the introduction of a finite photon mass.  The renormalization prescription is described in the previous section.
 
 The IR divergences  from the virtual diagrams are cancelled by real photon emission contributions, $H\to WW\gamma$. 
Due to the complex Lorentz structures of the SMEFT vertices, the calculation of the width $H\to WW\gamma$ through direct integration of the phase space is extremely intricate.
In order to calculate the real corrections we used the method developed in  \cite{Anastasiou:2002yz}, where the integration over the phase space is replaced with a loop integration.   
This is possible after we recognize that the Cutkosky rules allow us to replace the delta functions inside the phase space integrations with propagators:
\bea
2i\pi\delta(p^2-m^2)=\frac1{p^2-m^2+i0}-\frac1{p^2-m^2-i0}.
\label{eq:Cutkosky}
\eea
After making this replacement, we can treat the momenta of the outgoing particles as internal loop momenta, and the integration over the phase space becomes an integration over the loop momenta.
This allow us to use the IBP relations  to reduce the loop integrals in terms of Master Integrals (MI).  The methodology of this approach is described in Ref.  \cite{Anastasiou:2002yz}.

\begin{figure}[t]
\includegraphics[width=\textwidth]{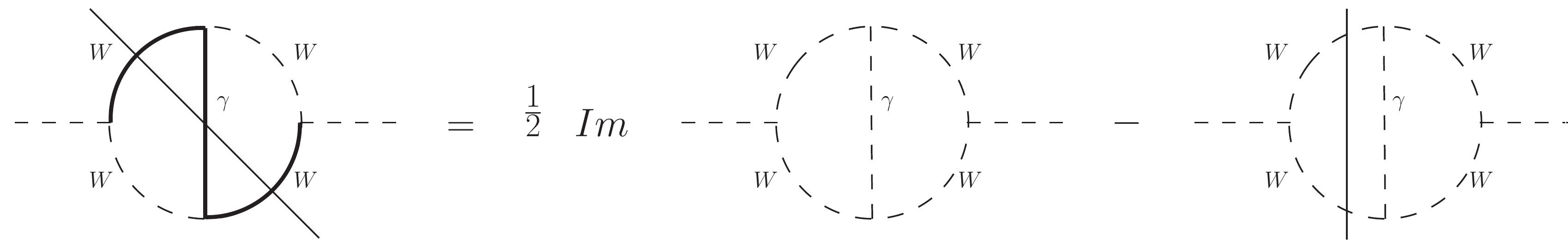}
\caption{Example of the calculation of $WW\gamma$ phase space. From the reduction we obtain the central integral. There are four possible ways to cut it: two over $WW$ and two over $WW\gamma$. Since we are interested in calculating the integral with a single physical cut over $WW\gamma$ (left integral), we need to subtract a counterterm (right integral).}
\label{fig:PhaseSpace}
\end{figure}

In the specific case of $H\to WW\gamma$, the integrals obtained are 2-point 2-loop integrals, for which a generic basis of MI is known  \cite{Tarasov:1997kx, Martin:2003qz}. The reduction was done using FIRE \cite{Smirnov:2014hma}.
Since many 2-point 2-loop MI are known analytically,  and the rest can be calculated numerically with high precision, for example using TSIL  \cite{Martin:2005qm}, we evaluate the MI directly and take the imaginary part of the result. 
An important caveat is that after the reduction  to MI, we have to select only the MI that still have a physical  $WW\gamma$ cut, while we  put to zero those that have lost one or more of the propagators generated by Eq.~(\ref{eq:Cutkosky}). An interesting consequence of this is that if an integral can be cut in more than one way it is necessary to add a  counterterm to cancel the extra imaginary part. As an example of this proceedure, see Fig.~\ref{fig:PhaseSpace}.

We have verified analytically that the IR divergences proportional to the photon mass cancel using this technique. 
 
The total width is then the sum of the virtual and real contributions, 
and  is given for $M_H=200~GeV$ by $\Gamma=\Gamma_0+\delta \Gamma_{NLO}$ with, 
\begin{eqnarray}
\delta\Gamma_{NLO}&=&
0.065253 +\biggl({1~TeV\over\Lambda}\biggr)^2\biggl\{\biggl[
   \biggl(
  190.1  - 70.52  X_\Lambda\biggr)  \C_{\phi\square}(\Lambda)\nonumber \\ &&
  +\biggl( -203.1  + 6.668   X_\Lambda\biggr)  \C_{\phi l}^{(3)}(\Lambda)
  +
  \biggl( -44.44  + 
   16.82 X_\Lambda\biggr)  \C_{\phi D}(\Lambda)
   \nonumber \\ &&
   +  \biggl( 
  -241.4 + 44.54 X_\Lambda \biggr)  \C_{\phi W}(\Lambda)
  +\biggl( 
  71.80  - 3.291  X_\Lambda\biggr)   \C_{ll}(\Lambda)\biggr]
\nonumber \\ && 
 +\biggl( 
  52.06   - 30.64 X_\Lambda\biggr)  \C_{uW}(\Lambda)
    + \biggl( -101.0   + 50.69  X_\Lambda\biggr) \C_{\phi q}^{(3)}(\Lambda)\nonumber \\ &&
    -\biggl( 5.191  +  32.85  X_\Lambda\biggr) \C_W(\Lambda)
    + \C_{lq}^{(3)}(\Lambda) \biggl( 
  18.54   - 23.72 X_\Lambda\biggr)
\nonumber \\ &&
  -8.434 \C_\phi(\Lambda)
  -1.157   \C_{u\phi}(\Lambda)
  +\biggl( 
   -0.9828 + 2.192 X_\Lambda \biggr) \C_{\phi B} (\Lambda)
 \nonumber \\ && 
+ \biggl( 
 - 17.24 + 3.256 X_\Lambda\biggr)   \C_{\phi WB}(\Lambda)
  -1.290  \C_{\phi l}^{(1)}(\Lambda)  \biggr\}  \times 10^{-4} \, ,
\end{eqnarray}
where the coefficients are evaluated at the scale $\Lambda$ and $X_\Lambda=\log(\Lambda^2/M_Z^2)$. (The terms in the square brackets
occur at tree level.)

We define the (on-shell) scaling factor at one-loop for $M_H=200~GeV$ and $\Lambda=1~TeV$,
 \begin{eqnarray}
 \mu_1(H\rightarrow WW)&=&{\Gamma_0+\delta \Gamma_{NLO}\over(\Gamma_0+\delta \Gamma_{NLO} )\mid_{SM} }\nonumber \\
&=&
1+\biggl[
0.1007   \C_{\phi\square}(\Lambda)-0.1295 \C_{\phi l}^{(3)}(\Lambda) -0.02525  \C_{\phi D}(\Lambda)-0.2269  \C_{\phi W}(\Lambda)
\nonumber \\ &&
+0.06209  \C_{ll(\Lambda)}\biggr]+
 \biggl\{-85.64  \C_{uW}(\Lambda)+128.2 \C_{\phi q}^{(3)}(\Lambda)-146.9 \C_W(\Lambda) 
\nonumber \\ &&
-85.94 \C_{lq}^{(3)}
+- 7.617 \C_{\phi} (\Lambda)
-1.04493\C_{u\phi} (\Lambda)
+ 8.604 C_{\phi B}(\Lambda)\nonumber \\ &&
-1.474 \C_{\phi WB}(\Lambda)
  -1.165\C_{\phi l}^{(1)}(\Lambda) \biggr\} \times 10^{-4}\,.
  \label{eq:mu1nlo}
\end{eqnarray}
The change in the coefficients of operators that appear at tree level (in the square brackets in Eq. \ref{eq:mu1nlo})
 is typically a few percent, while a few of the operators that first appear at one-loop 
have sizable coefficients and could potentially be probed in $H\rightarrow W^+W-$ decays. 

\section{$H\rightarrow \gamma\gamma$}
\label{sec:ggsec}

As a by-product of our  calculation of $H\rightarrow ZZ$ and $H\rightarrow Z \gamma$ \cite{Dawson:2018pyl}, we obtain the SMEFT
result for $H\rightarrow \gamma^\mu(p_1) \gamma^\nu (p_2) $ at one-loop.  
Gauge invariance requires that the one-loop amplitude take the form, 
\bea
{\cal{A}}^{\mu\nu}&=& F
 \biggl(g^{\mu\nu}-\frac{ p_1^\nu p_2^\mu}{p_1\cdot p_2}\biggr)\nonumber \\
&=& \biggl(F^{0}_{SMEFT}+F^{1}_{SM}
+F^{1}_{SMEFT}\biggr)
 \biggl(g^{\mu\nu}-\frac{ p_1^\nu p_2^\mu}{p_1\cdot p_2}\biggr)\, ,
\label{eq:ampres}
\eea
where we have broken up the coefficient into the tree level SMEFT piece, $F^0_{SMEFT}$, the one-loop
SM piece, $F^1_{SM}$,  and the one loop SMEFT contribution, $F^1_{SMEFT}$.

Initially, we take $M_W$, $M_Z$ and $G_\mu$ as input parameters.  
At tree level, there is only the SMEFT contribution,
\begin{eqnarray}
F_{SMEFT}^0&=&-8M_H^2
 {M_W^2\over \Lambda^2}\sqrt{\sqrt{2}G_\mu}
\biggl(1-{M_W^2\over M_Z^2}\biggr) \C_{\gamma
\gamma}
\end{eqnarray}
where
\begin{equation}
\C_{\gamma \gamma}\equiv {1\over 4\sqrt{2}G_\mu M_W^2}
\biggl(
\C_{\phi W}+{M_W^2\over M_Z^2-M_W^2} \C_{\phi B}
-{M_W\over\sqrt{M_Z^2-M_W^2}} \C_{\phi  WB}\biggr)\, .
\label{eq:cggdef}
\end{equation}

The    analytic one-loop   SM result can be found in many places \cite{Ellis:1975ap}
and we write the results numerically. 
The SMEFT logarithms contributing to $F_{SMEFT}^1$ can be obtained from the anomalous dimensions given in 
Ref.~ \cite{Alonso:2013hga}
and are written in terms of
\begin{equation}
X_\Lambda=\log\biggl({\Lambda^2\over M_Z^2}\biggr)\, .
\end{equation}
The  full 1-loop SMEFT  result for $H\rightarrow \gamma\gamma$ is 
 extracted from the calculation of Ref. ~ \cite{Dawson:2018pyl}  
 (for the input parameters of Eq. \ref{eq:inputs}).  Our complete result is, 
\begin{itemize}
\item
$G_\mu,~ M_W,~M_Z$  input parameter scheme:
\end{itemize}
\begin{eqnarray}
F_{SMEFT}^0&=&\biggl({1~TeV\over \Lambda}\biggr)^2\biggl\{
-5.988 \C_{\phi B}(\Lambda) - 1.718 \C_{\phi W}(\Lambda) + 3.207 \C_{\phi WB}(\Lambda)\biggr\}
 \nonumber \\ 
 F_{SM}^1&=& 0.2483   \nonumber \\ 
F_{SMEFT}^1&=&\biggl({1~TeV\over \Lambda}\biggr)^2\biggl\{\biggl[
\biggl(0.3636   + 0.1336 X_\Lambda\biggr) \C_{\phi B}(\Lambda)+
 \biggl(0.02362  + 0.01456  X_\Lambda\biggr)\C_{\phi W}(\Lambda)
 \nonumber \\ &&
+ \biggl( -0.1272  - 0.06487 X_\Lambda\biggr)\C_{\phi WB}(\Lambda)\biggr]
 \nonumber \\ &&
+\biggl( 0.01304  - 0.02725  X_\Lambda\biggr)\C_W(\Lambda)
  +0.01505 \C_{\phi \square}(\Lambda)
    -0.03000 \C_{\phi D}(\Lambda)
  \nonumber \\ && 
 + 0.004279 \C_{u\phi}(\Lambda)
  + \biggl( 0.1276  - 0.05649 X_\Lambda\biggr)\C_{uW}(\Lambda)
  + \biggl(0.2383 - 0.1055 X_\Lambda\biggr)\C_{uB} (\Lambda)
 \nonumber \\ &&
 -0.04516  \C_{\phi l}^{(3)}(\Lambda)
+  0.02258 \C_{ll}(\Lambda) \, .
  \biggr\}
   \label{eq:resus}
 \end{eqnarray}
 The coefficients are given in $GeV$ and are evaluated at the scale $\Lambda$.  This is the appropriate scale for matching 
 with  high-scale UV complete models\cite{Brehmer:2015rna,Dawson:2017vgm,deBlas:2017xtg,delAguila:2016zcb}.
 However, it should be highlighted that in order to be consistent, the matching with the UV model should be computed at  NLO. 
 Moreover a more general calculation would require the computation of the RG evolution of the coefficients also at NLO order.
 This is particularly true when the separation between the electroweak and the EFT scales is very large and it becomes necessary to resum the logarithms $\log\biggl(\Lambda^2/M_Z^2\biggr)$.  Our calculation is a first step in this program.  
 Note the dependence at one-loop on coefficients that do not appear at tree level, leading to the interesting
 possibility of obtaining limits on previously unconstrained coefficients. 
 
 We recalculate the result using 
 $\alpha$, $G_\mu$,  $M_Z$,  and  $M_W$ as inputs, as described in Sec. \ref{sec:basics}. 
  For notational convenience, the amplitude is expressed as, 
 \bea
{\cal{A}}^{\mu\nu}&=&\alpha\biggl({\hat F}^{0}_{SMEFT}+{\hat F}^{1}_{SM}
+{\hat F}^{1}_{SMEFT}\biggr)
 \biggl(g^{\mu\nu}-\frac{ p_1^\nu p_2^\mu}{p_1\cdot p_2}\biggr)\, .
\label{eq:ampres}
\eea
 The result is, 
 \begin{itemize}
 \item
 {$\alpha,~ G_\mu,~M_Z, ~M_W$} input parameter scheme:
 \end{itemize}
 \begin{eqnarray}
\alpha {\hat F}_{SMEFT}^0&=&\biggl({1~TeV\over \Lambda}\biggr)^2\biggl\{-5.778  \C_{\phi B} (\Lambda)
 - 1.657  \C_{\phi W}(\Lambda) + 3.095 \C_{\phi WB}(\Lambda)\biggr\}
 \nonumber \\
\alpha {\hat F}_{SM}^1&=& 0.2396 
\nonumber \\
\alpha {\hat F}^1_{SMEFT}&=&\biggl({1~TeV\over \Lambda}\biggr)^2\biggl\{\biggl[
\biggr(  0.1234  + 0.1290X_\Lambda\biggr)\C_{\phi B}(\Lambda)+
 \biggr( -0.04246  + 0.01405 X_\Lambda\biggr)\C_{\phi W}(\Lambda)
 \nonumber \\
 &&+ \biggr( 0.05329  - 0.06260 X_\Lambda\biggr)\C_{\phi WB}(\Lambda)\biggr]\nonumber \\ &&
 +\biggr(  0.01259  - 0.02630 X_\Lambda\biggr)\C_W(\Lambda)
  +0.01452 \C_{\phi \square} (\Lambda)
   -0.003631  \C_{\phi D}(\Lambda)
  \nonumber \\ &&
 + 0.004129 \C_{u\phi}(\Lambda)
 +\biggr(  0.1231  - 0.05451 X_\Lambda\biggr)\C_{uW}(\Lambda)
  +\biggr(  0.2299  - 0.1018 X_\Lambda\biggr) \C_{uB}(\Lambda)
 \nonumber \\ &&
   -0.01452 \C_{\phi l}^{(3)}(\Lambda)
  + 0.007262 \C_{ll}(\Lambda)
\biggr\}\, .
\label{eq:alres}
 \end{eqnarray}
 We find that the coefficients calculated in the {$\alpha,~ G_\mu,~M_Z, ~M_W$} scheme are in agreement with those calculated in \cite{Dedes:2018seb}.
We also note that in both  input schemes, typically the
coefficients of the logarithms are of similar sizes to the constant pieces and that the differences are small in most cases.  
There are,  however, a few coefficients where the effect of the choice of the input parameter scheme is significant: comparing Eqs. (\ref{eq:resus}) and (\ref{eq:alres}) one can notice a factor $\sim 3$ between the coefficients of $(\C_{ll}-2\C_{\phi l}^{(3)})$ and a factor $\sim 8$ between the coefficients of $\C_{\phi D}$. In particular, while  $\C_{\phi\Box}$ and $\C_{\phi D}$ appear in Eq. (\ref{eq:alres}) in the combination $(\C_{\phi\Box}-\frac14\C_{\phi D})$ that is connected to the redefinition of the Higgs field in Eq. (\ref{eq:HiggsShift}), in Eq. (\ref{eq:resus}) one can verify that this simple relation is spoiled and the coefficients appear in the combination $(\C_{\phi\Box}-\frac{M_W^2+M_Z^2}{4(M_Z^2-M_W^2)}\C_{\phi D})\sim(\C_{\phi\Box}-2\C_{\phi D})$. These differences are due to the fact that using $\alpha,\, G_\mu,\, M_Z,\, M_W$ as input parameters modifies the counting of $\C_{\phi D}$ and $(\C_{ll}-2\C_{\phi l}^{(3)})$ that enter in the relation between the Lagrangian parameters and the input parameters $\alpha$ and $G_\mu$ respectively,  (see Eqs. 3.4-3.6 and 3.20 of \cite{Brivio:2017bnu}, for example).
Notice that the change of input parameter scheme affects also the coefficient of $\C_{\phi WB}$ that is present in the relation between the Lagrangian parameters and $\alpha$. However this effect is hidden in the relation between Eqs. (\ref{eq:resus}) and (\ref{eq:alres}) by the fact that $\C_{\phi WB}$ appears already at the LO with a large coefficient. The effect is instead evident in Eqs. (\ref{eq:resaaGF}) and (\ref{eq:resaaal}) in the appendix, where $\C_{\phi WB}$ appears only at the NLO.
The loop corrections to $C_{\phi W}$, $C_{\phi B}$ and $C_{\phi WB} $ are on the order of $1-2\%$,
relative to the tree level results.

\begin{figure}[t]
\includegraphics[width=0.4\textwidth]{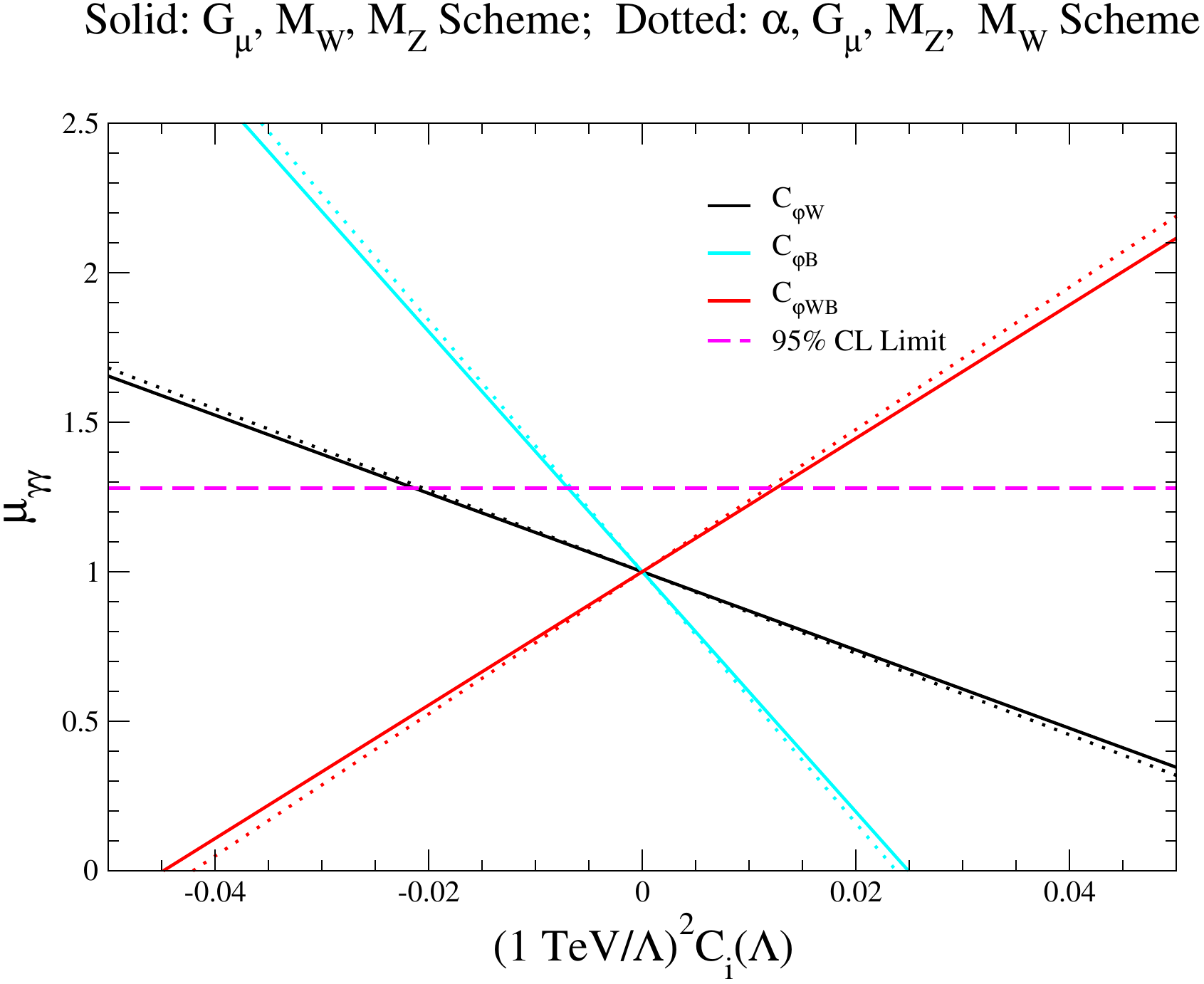}
\hspace*{0.1\textwidth}
\caption{Contributions to $\mu_{\gamma\gamma}$ when one operator at a time is varied, setting the remaining operators to $0$. The coefficients are evaluated at
the scale $\Lambda=1~TeV$.}
\label{fig:tree}
\end{figure}

We study the numerical consequences of our calculations by considering\footnote{Eq. \ref{eq:mudef} is in the $G_\mu, M_Z, M_W$ scheme.},
\begin{eqnarray}
\mu_{\gamma \gamma}&\equiv& {\Gamma(H\rightarrow\gamma\gamma)\over 
\Gamma(H\rightarrow \gamma\gamma)_{SM}}\nonumber \\
&=& 1+{2(F_{SMEFT}^0+F_{SMEFT}^1)\over F_{SM}^1}+{\cal{O}}\biggl({1\over\Lambda^4}\biggr)
\nonumber \\ &&=
1+\biggl[-40.15\C_{\phi B}(\Lambda)
-13.08\C_{\phi W}(\Lambda)
 +22.40 \C_{\phi WB}(\Lambda)\biggr]\nonumber \\ &&
 -0.9463\C_W(\Lambda)
  + 0.1212  \C_{\phi \square} (\Lambda)
   -0.2417  \C_{\phi D}(\Lambda)
   \nonumber \\ && 
   +0.03447 \C_{u\phi}(\Lambda)
 -1.151 \C_{uW}(\Lambda)
 -2.150\C_{uB}(\Lambda)
 \nonumber \\ &&
   -0.3637 \C_{\phi l}^{(3)}(\Lambda)
   + 0.1819 \C_{ll}(\Lambda)
\, .
\label{eq:mudef}
\end{eqnarray}

Our results can be compared with the limits from ATLAS \cite{Aaboud:2018xdt,Khachatryan:2016vau} and 
CMS \cite{Sirunyan:2018ouh,Khachatryan:2016vau},
\begin{eqnarray}
ATLAS, ~Run-2: & \mu_{\gamma\gamma}=& 0.99\pm 0.15\nonumber \\
ATLAS, ~Run-1: &  \mu_{\gamma\gamma}=&1.14 \pm 0.27 \nonumber \\
CMS,~Run-2: & \mu_{\gamma\gamma}=&1.18\pm 0.17\nonumber \\
CMS, ~Run-1: & \mu_{\gamma\gamma}=&1.11\pm 0.25\, .
\end{eqnarray}

We make the simplifying assumption that there are no cancellations between terms  and require that no single contribution
saturate the experimental bound.  This is probably a poor assumption, since in any specific model, there are relations between
the SMEFT coefficients \cite{Dawson:2017vgm,deBlas:2017xtg,Henning:2014wua}.  
When the complete set of one-loop SMEFT predictions to Higgs decay is known, it will be possible to do a global fit incorporating
these effects. 
In extracting the bounds, we are also ignoring the NLO effects induced by the matching with a UV theory, and the RG evolution discussed at the beginning of this session.  A full NLO RGE  calculation would be needed in order to reliably understand the size of these affects.   Our bounds are therefore only rough estimates of the sensitivity.
In Fig. \ref{fig:tree} we show the bounds on the coefficients that occur at tree level.  The argument of
the logarithms is evaluated at $\Lambda=1~TeV$.  The solid lines are the contributions in the $G_\mu,M_Z,M_W$ scheme and the dotted lines are the 
$G_\mu,M_Z,M_W,\alpha$  scheme.         Requiring
that $0 < \mu_{\gamma\gamma}< 1.28 $, we find for $\Lambda=1$ TeV,\footnote{Ref.   \cite{Dedes:2018seb} requires $0.85 < \mu_{\gamma\gamma}<1.15$ and so finds somewhat more restrictive
limits. The coefficients in Ref.  \cite{Dedes:2018seb}  are evaluated at the scale $\mu=M_W$.}
\begin{eqnarray}
&\mid \C_{\phi W}(\Lambda)\mid & <  0.02  \nonumber \\
  &\mid  \C_{\phi B}(\Lambda)\mid & <0.001    \qquad (H\rightarrow \gamma \gamma {\text{~limit}}) \nonumber \\
 &\mid \C_{\phi W B}(\Lambda)\mid & <0.01 \, \, .
\label{eq:cplims}
 \end{eqnarray}
 The coefficients can be evolved to low scales, $\mu\sim M_Z$, using the anomalous dimension matrix, 
 \begin{eqnarray}
 \C_i(M_Z)=\C_i(\Lambda)-{\gamma_{ij} \C_j\over 16\pi^2}\log\biggl({\Lambda\over M_Z }\biggr).
 \end{eqnarray}
 where the anomalous dimension matrices can be found in Refs.  \cite{Alonso:2013hga}  and the analogous numerical result to Eq. \ref{eq:mudef}, but with the coefficients
 evaluated at a low scale, can be found in Ref.   \cite{Dedes:2018seb}.

\begin{figure}[t]
\includegraphics[width=0.4\textwidth]{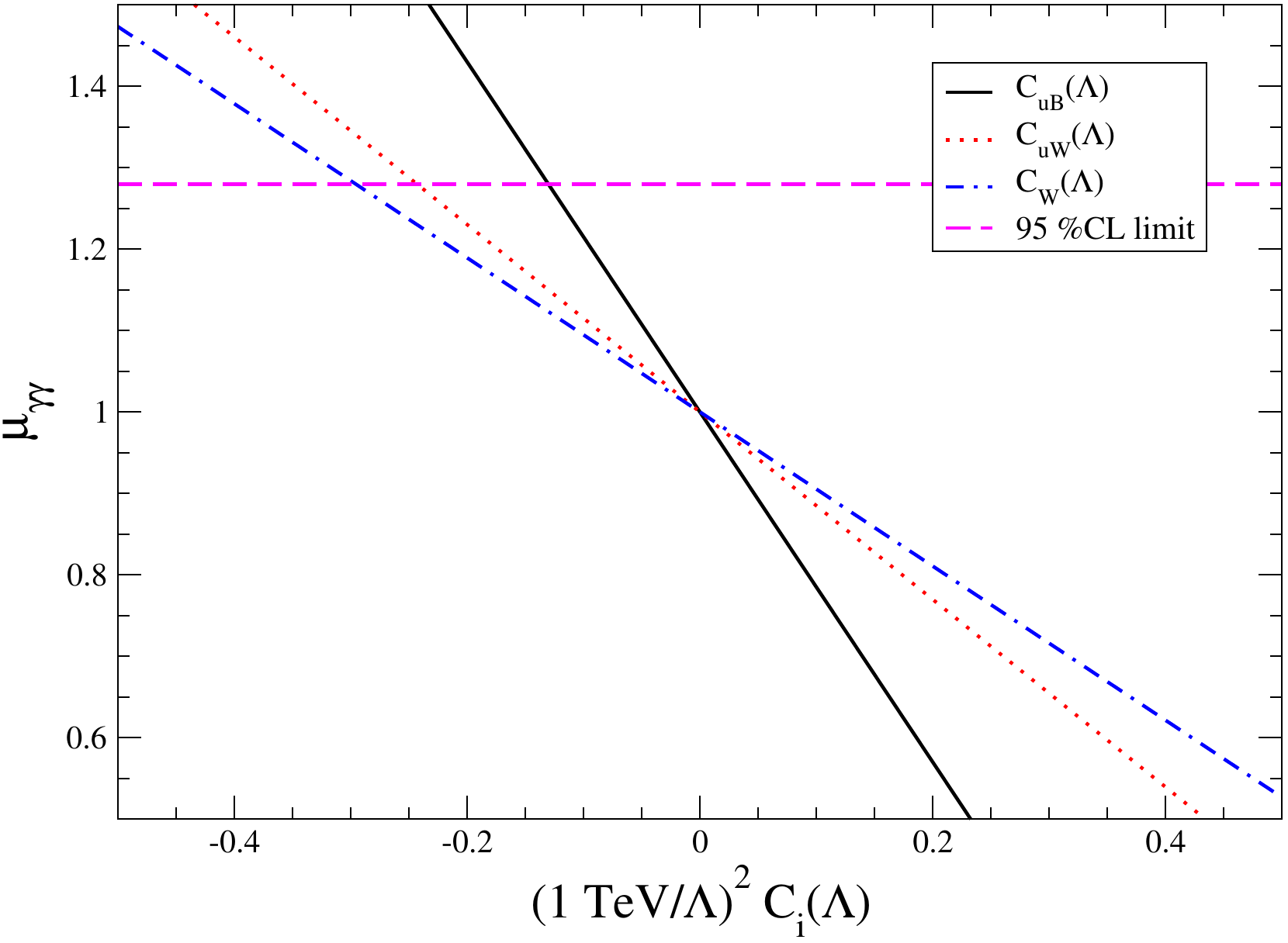}
\includegraphics[width=0.4\textwidth]{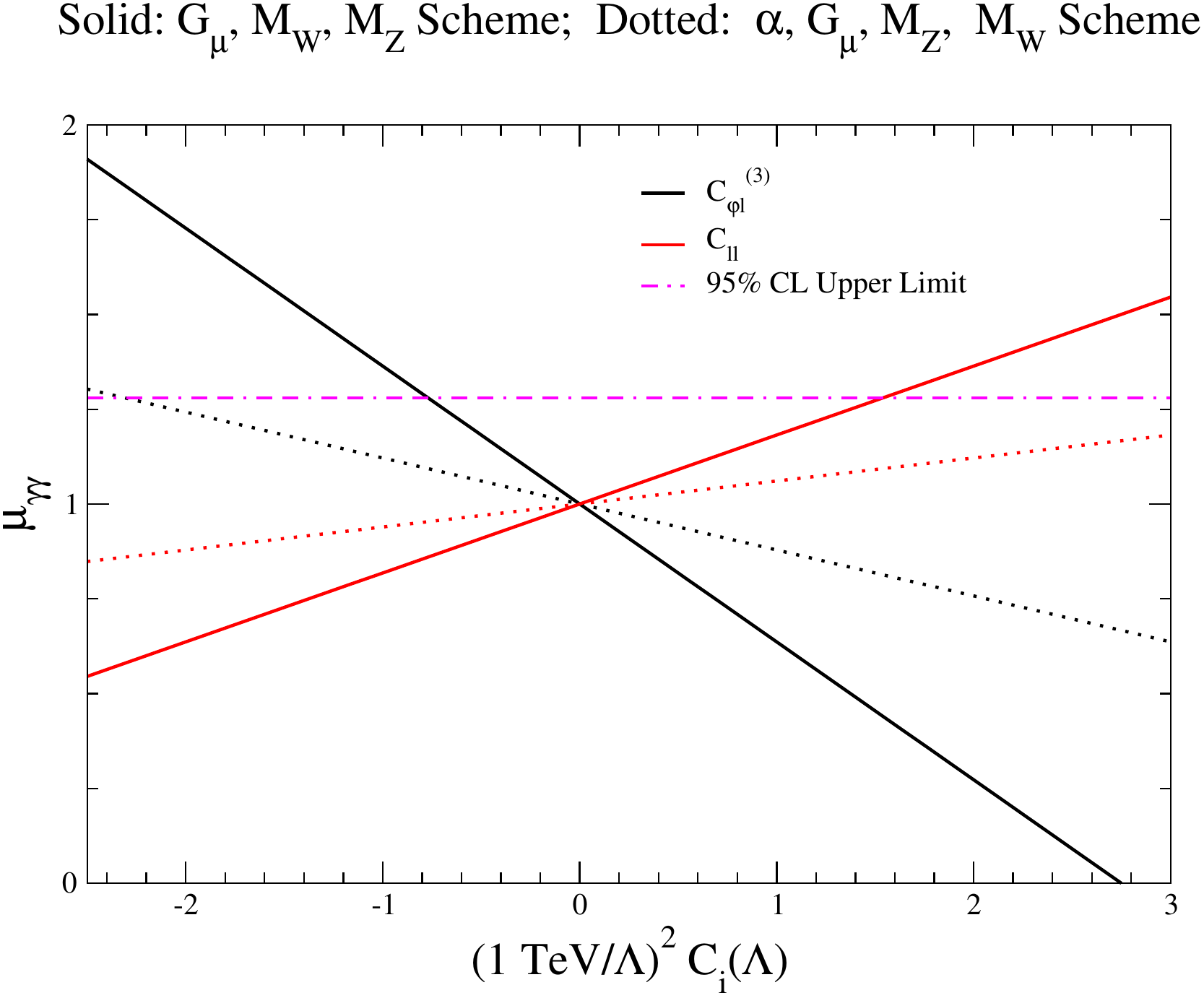}
\hspace*{0.1\textwidth}
\caption{Contributions to $\mu_{\gamma\gamma}$ when one operator at a time is varied, setting the remaining operators to $0$. The coefficients
are evaluated at the scale $\Lambda=1~TeV$.
On the LHS, the solid and dotted lines are indistinguishable. }
\label{fig:letcomp}
\end{figure}

On the LHS of Fig. \ref{fig:letcomp} we show the contributions to $\mu_{\gamma\gamma}$ from $\C_{uB}$ and $\C_{uW}$.  These coefficients first appear at loop
level and it is interesting that $H\rightarrow \gamma\gamma$ has the potential to place
limits on them.   We find (for $\Lambda=1$ TeV),
\begin{eqnarray}
 \mid \C_{uB}(\Lambda)\mid  &< &0.14  \qquad (H\rightarrow \gamma \gamma {\text{~limit}}) \nonumber \\
  \mid \C_{uW}(\Lambda)\mid  & <&  0.23 \, \, .
 \end{eqnarray}
 The contribution from $\C_W$ is shown on the LHS of Fig. \ref{fig:letcomp}.  This operator is particularly interesting 
 because it contributes to $W^+W^-$ pair production \cite{,Falkowski:2014tna,Baglio:2017bfe}.  
 Translating the tree level results of Ref.  \cite{Alves:2018nof} into our notation, we have for $\Lambda=1~TeV$,
 \begin{equation}
 |\C_W| <   0.08  \qquad (W^+W^-~{\text{limit}})\, .
 \end{equation}
 From Fig. \ref{fig:letcomp}, the limit on $\C_W$ from $H\rightarrow \gamma\gamma$ assuming that $\C_W$ is
 the only non-zero coefficient is $\mid \C_W\mid <0.3$, significantly weaker than the limit from gauge boson pair production.
 
 The contributions of $\C_{\phi l}^{(3)}$ and $\C_{ll}$ are particularly interesting because they contribute to $G_\mu$ at tree level and are
 shown on the RHS of Fig. \ref{fig:letcomp}.
 From Eq. \ref{eq:gdef}, they always contribute in the combination $\C_{\phi l}^{(3)}-{1\over 2}\C_{ll}$.  The over-all numerical factor between the schemes is just the difference in input parameters.

\section{Conclusions} 
\label{sec:conc}

We have computed the one-loop electroweak corrections to the decays $H\rightarrow \gamma\gamma$ and $H\rightarrow W^+W^-$
in the SMEFT.  The results are presented in simple forms, useful for comparison with LHC data and for matching with the predictions of
UV complete theories.

The decay $H\rightarrow \gamma\gamma$ is found using two different  input parameter schemes and the numerical dependence on the scheme choice
is negligible except for the coefficients of the operators $O_{\phi l}^{(3)}$ and $O_{ll}$ that contribute to $G_\mu$ at tree level.  For operators that contribute
to $H\rightarrow \gamma\gamma$  at tree level, the effect of the NLO corrections is a few percent.  However, the NLO result offers the possibility of constraining operators that first appear at one-loop.

The  real corrections to the on-shell decay $H\rightarrow W^+W^- \gamma$  are determined by transforming the 3 body final state phase space into 2 loop integrals,
while the virtual corrections are obtained using standard techniques.  The relatively large size of some of the one-loop contributions suggests that a complete
calculation of the off-shell decay $H\rightarrow $ 4 fermions at one-loop in the SMEFT is of interest.

\section*{Acknowledgements}
We thank Ahmed Ismail for discussions. We thank Athanasios Dedes and  Michalis Paraskevas for comments on the first version this paper.
S.D.   and P.P.G are supported by the U.S. Department of Energy under Grant Contract  de-sc0012704.

\section{Appendix}

It is interesting to  replace $\C_{\phi B}$ with  $\C_{\gamma \gamma}$.  Note that the tree level relation of Eq. \ref{eq:cggdef} cannot be used, but
we need the full one-loop calculation for consistency.      The results in $GeV$ with $\Lambda=1~TeV$ are, 
\begin{itemize}
\item
$G_\mu, ~M_W,~M_Z$ input parameter scheme:
\end{itemize}
\begin{eqnarray}
F_{SMEFT}^0&=& - 0.73  \biggl({1~TeV\over \Lambda}\biggr)^2  \C_{\gamma \gamma}(\Lambda)
\nonumber \\ 
F_{SM}^1&=&  0.2483  \nonumber \\ 
F_{SMEFT}^1&=& \biggl({1~TeV\over \Lambda}\biggr)^2   \biggl\{
 \biggl( 0.009007  + 0.01247  X_\Lambda\biggr) \C_{\gamma\gamma}(\Lambda)
-  \biggl(0.01904  +0.01549 X_\Lambda\biggr)  \C_{\phi WB} (\Lambda)
  \nonumber \\ &&
 +
\biggl( 0.01304  - 0.02725 X_\Lambda\biggr)  \C_W(\Lambda)
  + 0.01505  \C_{\phi\square}(\Lambda) 
  -0.03000   \C_{\phi D}(\Lambda)
+ 0.004279\C_ {u\phi}  (\Lambda)
\nonumber \\ &&
 + 0.01203  \C_{\phi W} (\Lambda)
+\biggl( 0.1276  - 0.05649  X_\Lambda\biggr) \C_{uW}(\Lambda)
 +\biggl( 0.2383  - 0.1055  X_\Lambda\biggr) \C_{uB}(\Lambda)
 \nonumber \\ &&
 -0.04516 \C_{\phi l}^{(3)}(\Lambda)
   +0.02258 \C_{ll}(\Lambda)
      \biggr\}
      \label{eq:resaaGF}
 \end{eqnarray}

\begin{itemize}
 \item
 {$\alpha, ~G_\mu,~M_Z,~M_W$} input parameter scheme:
 \end{itemize}
 The coefficients are, 
 \begin{eqnarray}
 \alpha{\hat F}_{SMEFT}^0&=& - \biggl({1~TeV\over \Lambda}\biggr)^2 
 0.7066  
\C_{\gamma \gamma} (\Lambda)\nonumber \\
\alpha {\hat F}_{SM}^1&=& 0.2396 
\nonumber \\
\alpha{\hat F}_{SMEFT}^1&=& \biggl({1~TeV\over \Lambda}\biggr)^2  \biggl\{
 \biggl(-0.01913  + 0.01203  X_\Lambda\biggr) \C_{\gamma \gamma} (\Lambda)
+\biggl(0.03587  - 0.01495   X_\Lambda\biggr)\C_{\phi W B} (\Lambda)
\nonumber \\ &&
 +\biggl(0.01259  - 0.02630   X_\Lambda\biggr) \C_W(\Lambda)
 + 0.01452  \C_{\phi\square}(\Lambda)
-0.003631 \C_{\phi D} (\Lambda)
 +0.004129 \C_{u\phi} (\Lambda)
 \nonumber \\ && 
 + 0.01161  \C_{\phi W} (\Lambda)
+ \biggl( 0.1231  - 0.05451   X_\Lambda\biggr) \C_{uW}(\Lambda)
 +\biggl( 0.2299  - 0.1018   X_\Lambda\biggr) \C_{uB}(\Lambda)
 \nonumber \\ &&
   -0.01452  \C_{\phi l}^{(3)} (\Lambda)
+  0.007262   \C_{ll}(\Lambda)
   \biggr\}
   \label{eq:resaaal}
\end{eqnarray}

\bibliographystyle{utphys}
\bibliography{hgg_paper}

\end{document}